# An API for Development of User-Defined Scheduling Algorithms in Aneka PaaS Cloud Software: User Defined Schedulers in Aneka PaaS Cloud Software

*Rajinder Sandhu[a,b], Adel Nadjaran Toosi[c], and Rajkumar Buyya[a]*

*[a] CLOUDS Laboratory, School of Computing and Information Systems, University of Melbourne, Australia.*

*[b] Department of Computer Science and Engineering, Jaypee University of Information Technology, Waknaghat, India.*

*[c] Faculty of Information Technology, Monash University, Australia.*

## Abstract:
Cloud computing has been developed as one of the prominent paradigm for providing on demand resources to the end user based on signed service level agreement and pay as use model. Cloud computing provides resources using multitenant architecture where infrastructure is generated from multiple or single geographical distributed cloud datacenters. Scheduling of cloud application requests to cloud infrastructure is one of the main research area in cloud computing. Researchers have developed many scheduling applications for which they have used different simulators available in the market such as CloudSim. Performance of any scheduling algorithm will be different when applied to real time cloud environment as compared to simulation software. Aneka is one of the prominent PaaS software which allows users to develop cloud application using various programming models and underline infrastructure. In this chapter, a scheduling API is developed over the Aneka software platform which can be easily integrated with the Aneka software. Users can develop their own scheduling algorithms using this API and integrate it with Aneka software so that they can test their scheduling algorithm in real cloud environment. The proposed API provides all the required functionalities to integrate and schedule private, public or hybrid cloud with the Aneka software.

## 1. Introduction
Cloud computing has proved to be the most revolutionary technology of the last decade which resulted in many organizations moving toward cloud-based infrastructure [1]. From mobile applications to large data intensive applications are using cloud-based infrastructure for fulfilling their IT resource requirement. With more data generation, the need for cloud computing is increasing day by day for many emerging IT technologies such as Internet of Things and Big Data [2]. Cloud computing deployment models can be broadly classified into Public cloud, Private cloud and Hybrid cloud. Among all these models, hybrid cloud is gaining popularity with its features like infinite resources and cost benefits. Hybrid cloud utilizes public cloud resources if private cloud resources cannot complete the task with given Quality of Service (QoS) parameters. This makes hybrid cloud model a good candidate for many applications such as mobile devices, small industries, and other smart environments [3].

As cloud computing contains a colossal number of IT resources whether it is a private cloud or public cloud, it is difficult to test new algorithms for better and efficient scheduling. Many researchers use simulation tools to test and deploy different scheduling algorithms for different kind of applications in the cloud environment. A common used simulation tool is CloudSim [4] while other tools such as iFogSim [5] and IoTSim [6] are also gaining popularity for testing IoT based applications on cloud computing environments. But, results from even the most efficient simulation software always differ from actual results because many other aspects such as network, bandwidth also play an important role. Infrastructure as a Service (IaaS) provider such as Amazon EC2 [7], or Microsoft Azure [8] give access to underlying IT resources to the user. But these IaaS providers do not give rights to the end user for making any change in the scheduling policies for the application. There are many options in the market for PaaS where the end user can create individual tasks and submit them to the PaaS provider. Many PaaS providers do not provide access to underlying IT infrastructure making it very difficult to change and test new scheduling policies [9]. Due to these constraints, research in the development of new scheduling policies is taking a big hit for real cloud computing environments.

Aneka [10] is a PaaS cloud provider developed in Microsoft .net for developing cloud computing infrastructure and applications using various programming models and available infrastructure. Aneka supports programming models such as Task, Thread and Map-Reduce while users can develop their own model. In Aneka, infrastructure can be developed using a cluster of multicore machines, private cloud and public cloud. Aneka contains inbuilt scheduling policies which are used to schedule jobs on private cloud created using multicore machines or using private cloud software such as OpenStack. Aneka provides dynamic provisioning feature which allows Aneka applications to use public cloud such as Amazon EC2 and Microsoft Azure when desired QoS cannot be achieved using the private cloud setup [11]. Aneka is one of its kind which gives the end user full freedom to develop applications in many languages and deploy it on any infrastructure available with them. It gives full access and right of underlying infrastructure to end user and well as full access to SDK for development of applications [12]. Aneka has many features but it is difficult for the end user to create their custom scheduling and provisioning policies.

In this chapter, we propose to develop an API for Aneka which allows end users to create and integrate their custom scheduling algorithms with Aneka. The user can develop their own applications in Aneka and then create a customized scheduling policy according to their application needs. Aneka also provides user with inbuilt sample applications such as Mandelbrot, Image Convolution or Blast to test their new scheduling algorithm. The proposed API bridges the gap between access to scheduling algorithms and using real cloud setup to test and create them.

The rest of the chapter has been organized as follows. Section 2 provides the introduction about the architecture of Aneka. Section 3 explains the proposed API with all classes. Section 4 provides sample codes and related discussion. Section 5 provides performance evaluation. Finally, Section 6 concludes the chapter and provides future directions.

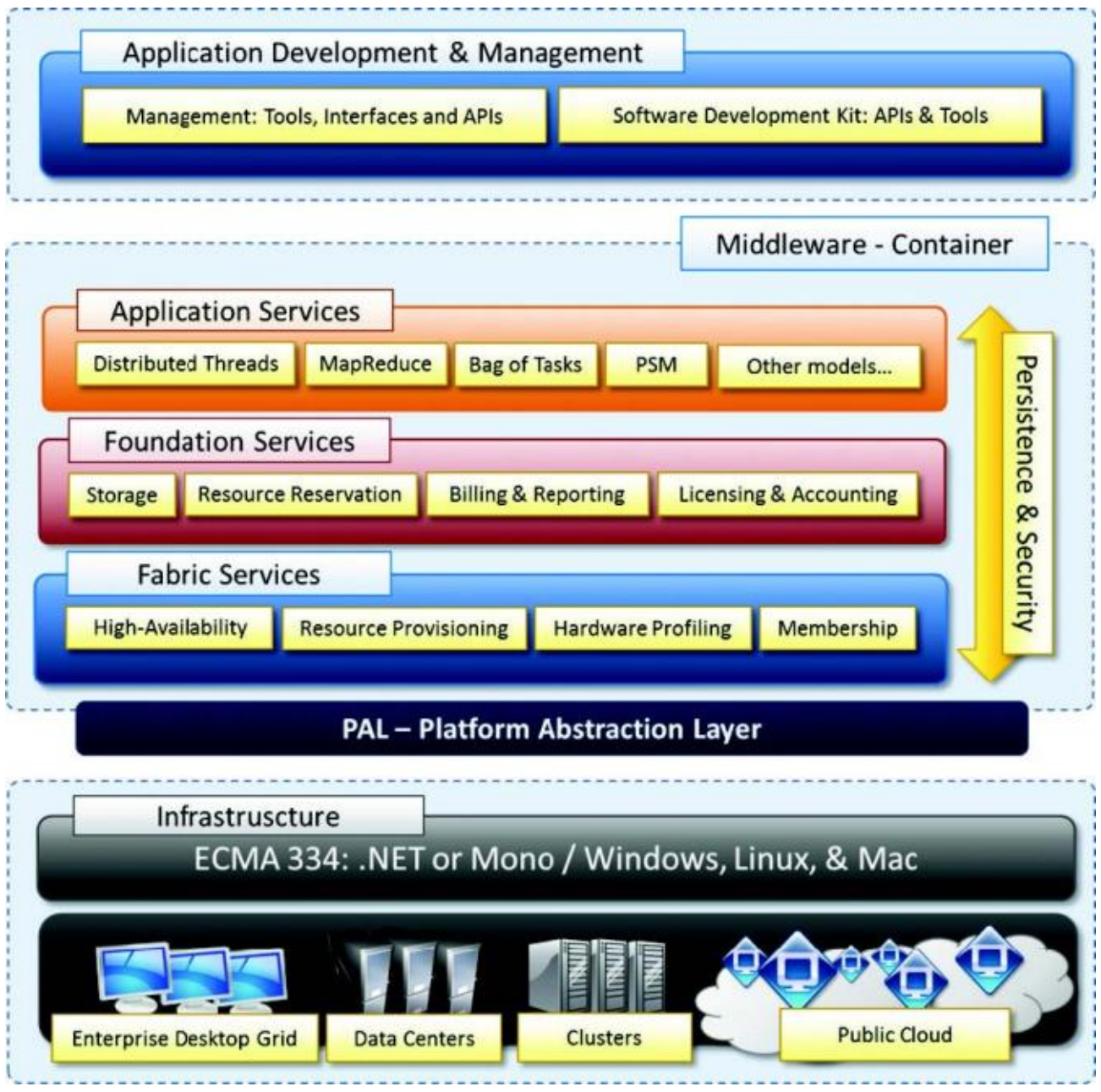

Figure 1. Overview of Aneka Architecture [11]

## 2. Aneka

Aneka is a PaaS cloud software that facilities the development and deployment of applications with underline support of .net framework. Figure 1 shows an overview of the architecture of Aneka. It contains three layers which are infrastructure, middleware and application development which allows the end user to change underline infrastructure and middleware to support rapid and customized development of cloud applications. The infrastructure of Aneka can be multicore machine, grid environment, cluster of machines, private cloud and public cloud. Public cloud is only required when scheduler decides that desired QoS cannot be achieved from other available resources. Middleware provides the support for programming models such as thread, task and Map-Reduce. It also provides the billing, accounting, resource reservation, hardware profiling and other services. These service in Aneka can be plugged and unplugged as and when required from the Management console provided in the application development and management layer. Aneka also contains many components which are discussed below:

- **Aneka Master Container:** Aneka master is mainly responsible for the scheduling and monitoring of application tasks and resources in the Aneka. Aneka master also manages the billing and reporting services of all the worker nodes attached with an Aneka master. The end user's application sends the task to Aneka master which based on the availability of worker nodes further schedules these tasks. If any worker nodes fail without executing the task, master reschedules the task o another available worker node.
- **Aneka Worker Container:** Aneka worker is the component which basically deals with the execution of the tasks. It contains the executor for each type of programming models available with the Aneka. After completing any assigned task, it sends the result back to master for compiling and starts waiting for new tasks. Worker node is operating system independent, it can be run on Linux or Windows machines.
- **Aneka Daemon:** Aneka daemons are the basic services which need to be installed in all the machines before Aneka master and worker can be installed.
- **Management Studio:** Management studio is the interactive interface which end user uses to create and manage cloud environments created inside the Aneka cloud. The end user can easily add private and public cloud resources, manages all added resources, create bills, monitor added resources, add file repositories, and check current statistics of the Aneka cloud.
- **Aneka SDK:** Aneka also provides SDK for the development of applications which can be directly deployed on the Aneka cloud created using management studio. These SDKs contains programming language specific libraries and other tools.

## 3. Proposed API

Figure 2 shows the proposed API for Aneka cloud which acts as an independent interface which user can use to change the scheduling policies of Aneka. The user level scheduler is the new or proposed scheduler by the end user for its specific application or in general use. This user level scheduler is integrated with Aneka using the proposed scheduling API. This section explains different classes and interfaces used in the proposed API so that end user can easily write their own scheduling policies.

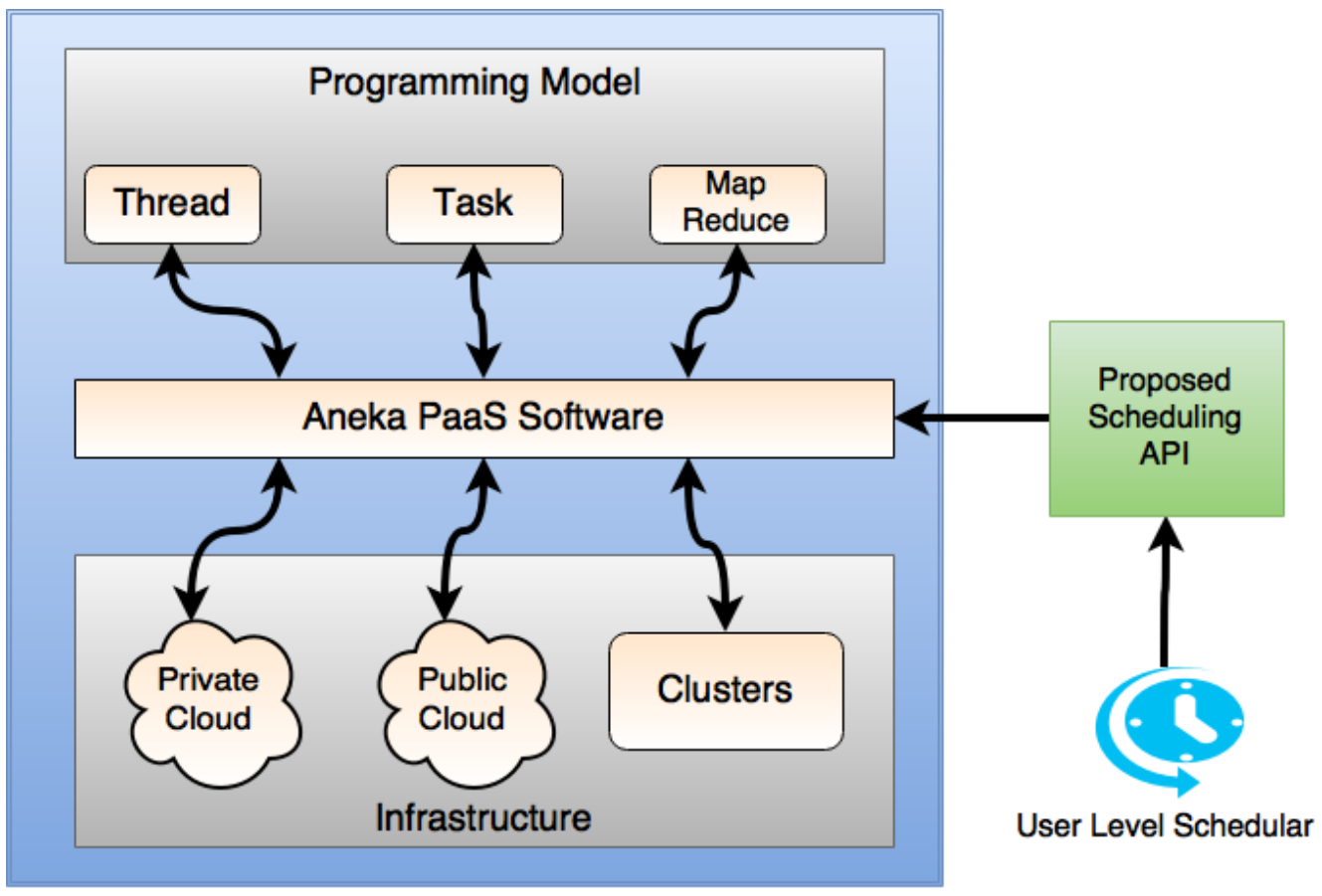

Figure 2. Framework for Scheduling API with Aneka PaaS Software

Aneka.Scheduling is the main project of proposed API which directly interacts with Aneka.Runtime project of Aneka to state which scheduling policy should be followed when any task arrives. Aneka.Runtime consults Aneka.Scheduling and based on the selected algorithm at the time of creation of Aneka master it schedules tasks on different worker nodes. Aneka.scheduling API contains six sub-projects out of which Aneka.Scheduling.Service and Aneka.Scheduling.Utils directly interacts with Aneka.Runtime project. These sub-projects are explained in details later in this chapter.

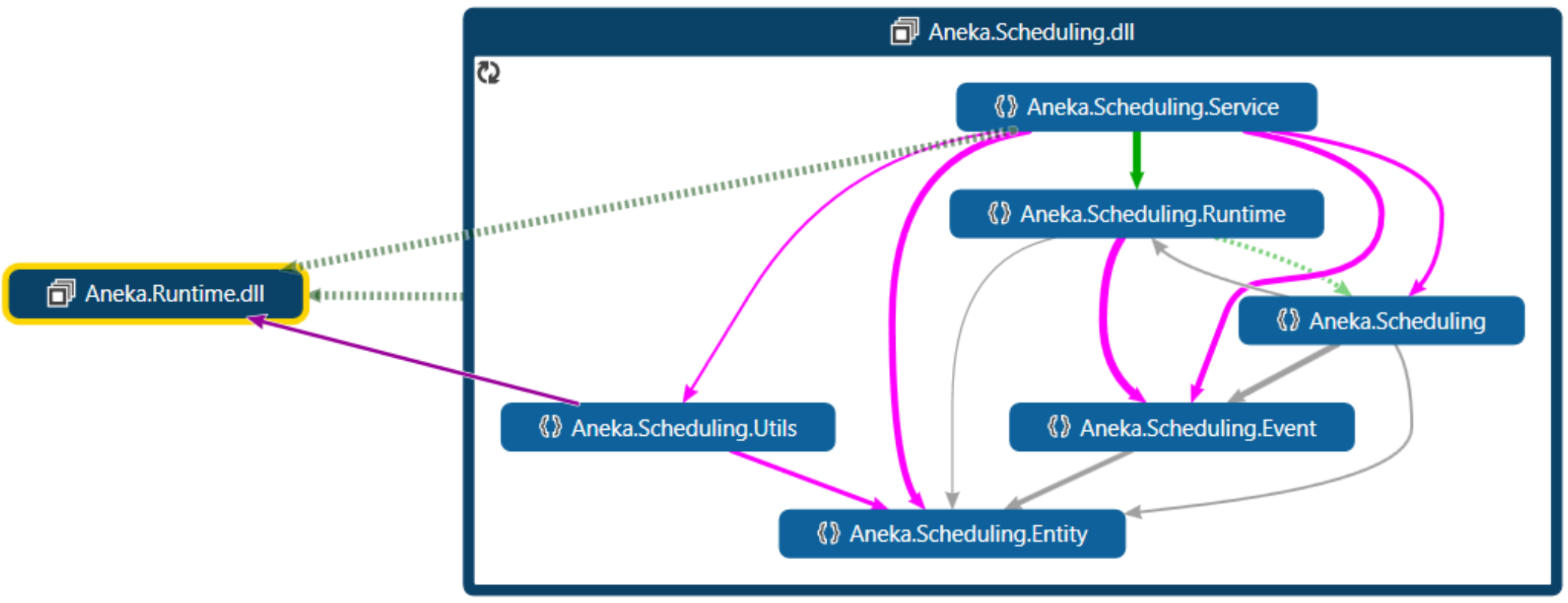

Figure 3: Code Map of proposed Aneka.Scheduling project in Aneka code

Aneka.Scheduling implements two interfaces which are ISchedulerContext and ISchedulerAlgorihtm as shown in Figure 4. Various events are associated with these interfaces which are triggered when there is a state change for the application or a task. Table 1 shows different event associated with Aneka.Scheduling.

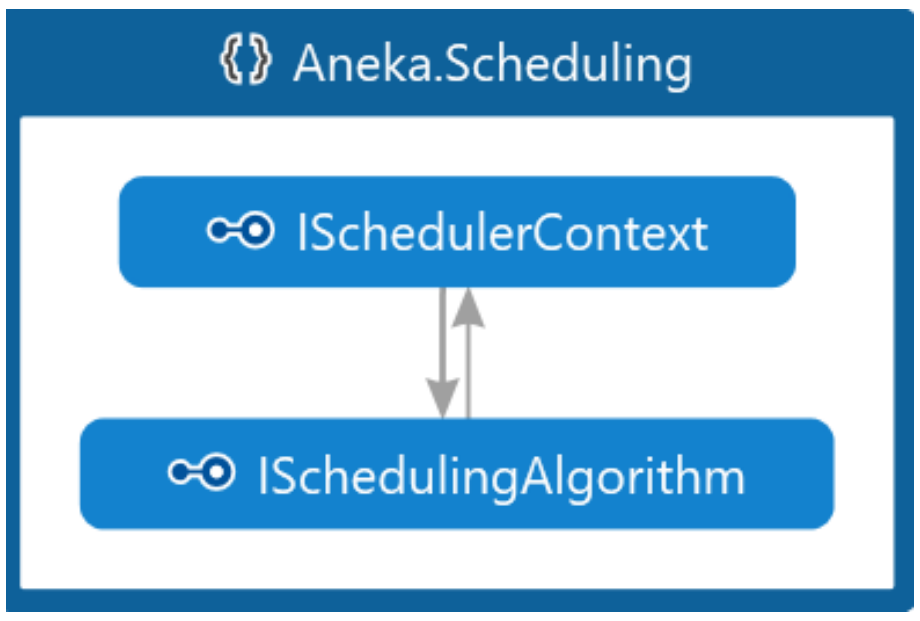

Figure 4 Interfaces implemented by Aneka.Scheduling Project

Table 1. Different Events associated with Aneka.Scheduling

| S. No. | Event | Description |
|---|---|---|
| 1. | SchedulerAlgorihtm | It selects the scheduling algorithm to use for scheduling tasks on worker nodes. |
| 2. | ResourceDisconnected | Event when a resource is disconnected. For every task assigned to this Resource the TaskFailed event will also be fired. |
| 3. | ResourceReconnected | Event when a resource is reconnected. |
| 4. | ResourceProvisionProcessed | Event when a resource provisioning request is processed. |
| 5. | ResourceProvisionRequested | Event when a resource provision request is triggered. |
| 6. | ResourceReleaseRequested | Event when a resource release request is triggered. |
| 7. | ResourcePoolsQueryRequested | This event is triggered when end user generates any resource pool related query. It is important event in case there are multiple pool in dynamic provisioning of Aneka cloud. |
| 8. | TaskFinished | Event when a task is finished. |
| 9. | TaskFailed | Event when a task is failed due the task failure. It is caused by something other than the resource disconnection |
| 10. | TaskAborted | Event when a task is aborted due to the user action. |
| 11. | TaskRequeued | Event when a task is requeued due the the user action |

### 3.1 Aneka.Scheduling.Runtime

It contains two classes and one interface as shown in the code map in Figure 5. This is responsible for decision making during runtime of any scheduling algorithm such as timings, resources available, task completed etc. SchedulerContextBase class registers all context-based activity happening in the scheduler to their specific events. This class also generates the exception and records it in the logger file so that end user can analyse the errors generated.

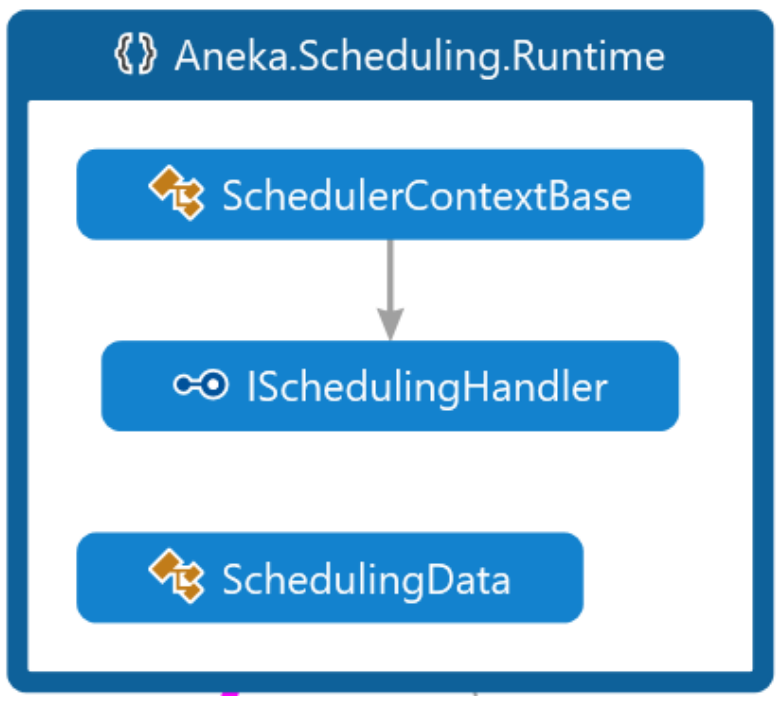

Figure 5 Classes and Interface of Aneka.Scheduling.Runtime

Major activities performed by this class are

- assign the value to the scheduling algorithm
- register the event handler for the forwarded assign work unit event from the scheduling algorithm
- hook the event handler for provision resources request
- hook the event handler for release resources request
- hook the event handler for a query for resource pools request
- register the scheduler context to the scheduling algorithm

Class SchedulingData extends the Aneka reporting data class which can capture the information about the scheduling data (mostly timing) of the specific task allocation. Various data collected by Scheduling data class are:

- **Queue Time:** A value representing the total waiting time of the task in seconds. This time is computed since the task is moved from queued to scheduled state.
- **Execution Time:** A value representing the total execution time of the task in seconds as seen by the scheduler service.
- **Task final State:** It provides the final state of the task submitted to the scheduler.

If the end user wants to add any new data point to the scheduling algorithm matrices, it can be easily added in this class. Interface ISchedulingHandler extends the IServiceNameAware interface of Aneka and provides a set of methods for the SchedulerService to specialize the activities of the scheduler. This interface allows separating all those management aspects that are common to several programming models, which reside in the SchedulerService class, from the specific aspects related to a given programming model, which reside in the component implementing this interface.

### 3.2 Aneka.Scheduling.Service

It contains two classes which are ScheulerService and IndependentSchedulingService as shown in code map in Figure 6. This is responsible for major scheduling responsibilities based on the selected algorithms.

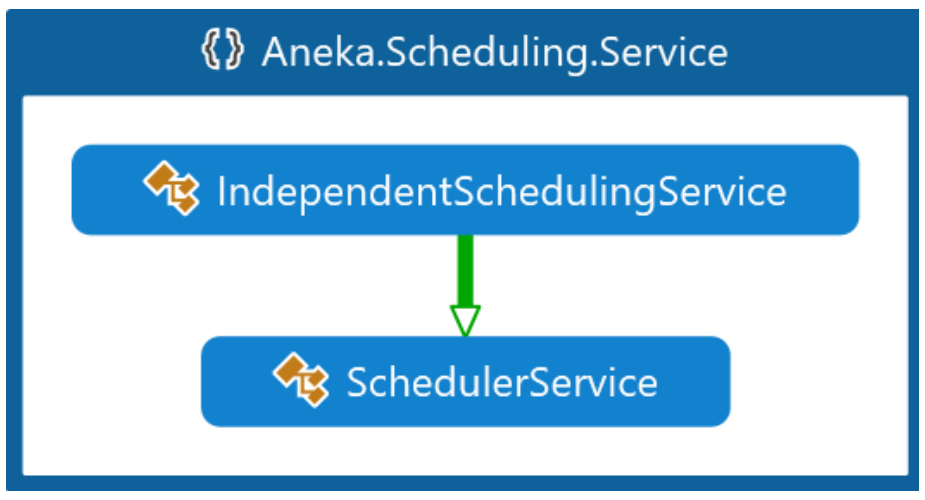

Figure 6 Code Map of Aneka.Scheduling.Service

Class SchedulerService specializes the ContextBase class of Aneka and implements the IService interface and IMembershipEventSink interface. It can be used as base application scheduler that needs to be further customized for handling the specific type of application according to the given programming model on which the application is based. The scheduled tasks directly performed by this scheduler are the interaction with the IApplicationStore interface to control the state of the application. It provides template methods that can be implemented by inherited classes to perform the WorkUnit level scheduled tasks. The activity of this service is supported by an implementation of ISchedulingHandler interface that deals directly with the programming model related scheduledTasks at a WorkUnit level.

Class IndependentSchedulingService defines a scheduling service for scheduling independent work units. Models featuring independent work units can use and specialize this scheduler policy in this class. Various parameters used in this class are:

- List of resources is maintained in this class on which scheduling can be done.
- Reservation list is present in this class.
- Starts and stops the scheduling service. Activates the scheduling algorithm and registers the resources with it.
- Fetches the list of WorkUnit instances that are in state Queued from the application store and delegates them back to the SchedulerAlgorithm.

### 3.3 Aneka.Scheduling.Algorithm

It contains different algorithm currently available with Aneka, all available scheduling algorithms extends AlgorihtmBase class for their proper execution as shown in Figure 7. This also contains one NewUserDefined which end user will use to code their own scheduling policy. End users can even extend or change already existing algorithms for more optimization. Class AlgorithmBase is the implementation class for the ISchedulingAlgorithm interface. This class can be used as a template for creating specialized algorithms because it provides the basic features for integrating scheduling algorithms into the Aneka scheduling service. Different variables used in this class are listed in Table 2 and methods available with AlgorithmBase class are listed in Table 3.

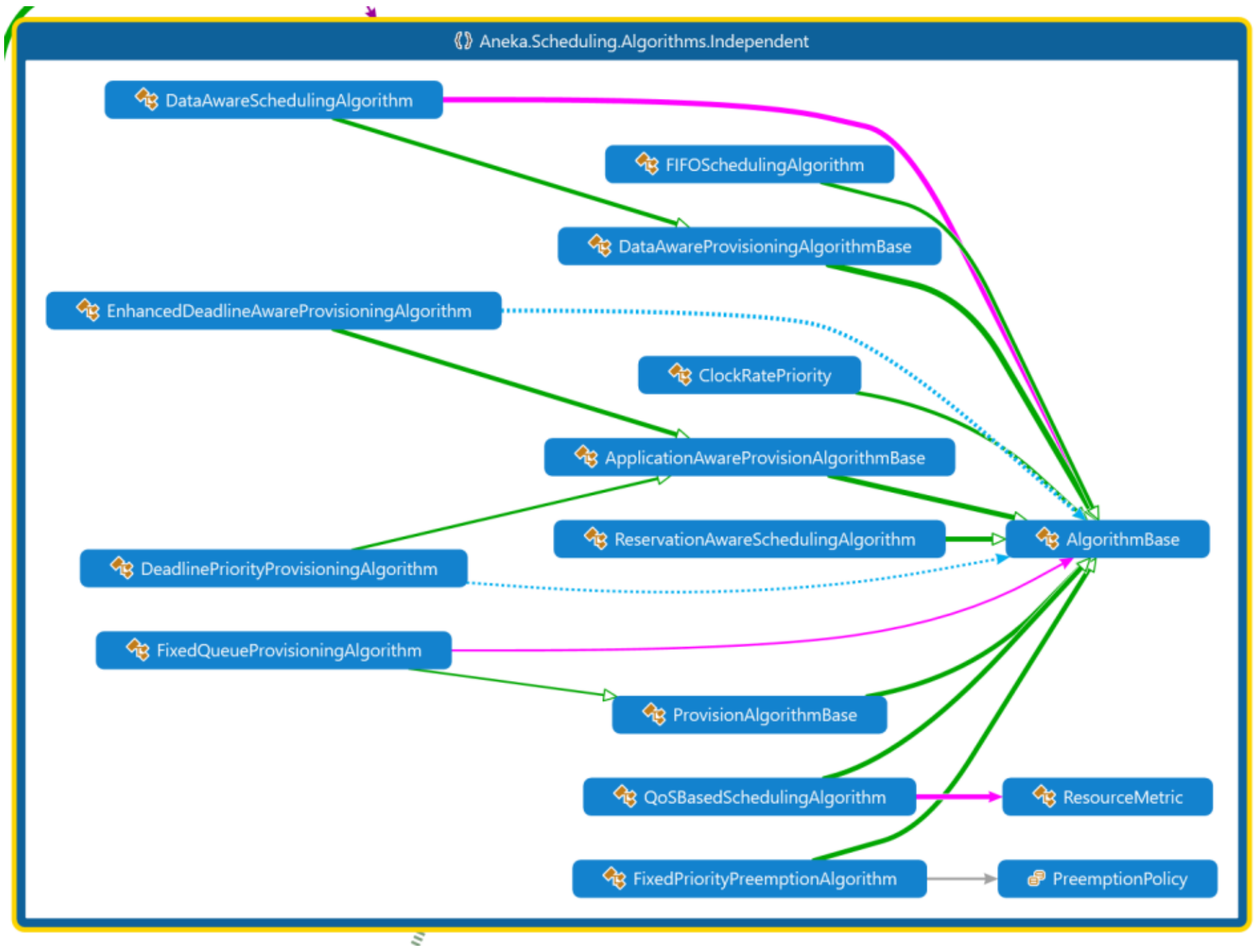

Figure 7 Code map of available scheduling algorithms

Table 2: Different variables/objects used in AlgorithmBase class

| S. No. | Variable/Objects | Parent Class | Description |
|---|---|---|---|
| 1. | rescheduledJobTimer | IDictionary | Dictionary mapping the each job reservationId to the timer used to reschedule them. |
| 2. | bKeepRunning | N.A. | While this variable is true the scheduling loop will bKeep running. |
| 3. | canSchedule | ManualResetEvent | Used to block the scheduling thread while there are no scheduledTasks or free resources. |
| 4. | scheduler | ISchedulerContext | Holds a reference to the context that the scheduler is interacting with. |
| 5. | SupportsProvisioning | N.A. | Whether the algorithm support dynamic provisioning or not |
| 6. | canFireEvent | AutoResetEvent | Used to block the event thread while there are no events to fire. |

Table 3: Methods Available in AlgorithmBase Class

| S. No. | Method Name | Description |
|---|---|---|
| 1. | Start | Start the scheduling loop for assigned tasks. |
| 2. | Stop | Stop the scheduling thread. |
| 3. | Schedule | This method is called repeatedly by the AlgorithBase class for making scheduling decisions. |
| 4. | AddTasks | Used to add new task to the scheduling queue. |
| 5. | GetNextTask | Used to get next task to schedule from the queue. |
| 6. | TaskFailed | Method is called when any task fails. Reschedule policy is checked in this method for the failed task. |
| 7. | TaskAborted | Method is called when user abort the assigned task. |
| 8. | TaskFinished | This is called when task is finished successfully and should be reported to the user its output values. |
| 9. | TaskRequeued | This method requeued the task if it is failed or aborted. |
| 10. | SetScheduler | This method hooks up the event handler for events fired with the desired scheduler. |
| 11. | StartScheduleTask | Start the process to schedule a given task to a given resource. |
| 12. | AddFreeResource | This method will be called when a resource has a task removed and it is not free. |
| 13. | HaveFreeResources | Return the value if there are free resource available that can be used. |
| 14. | RemoveFreeResource | Will be called when adding a task to the resource fills the slot of that resource. Can be called if resource get disconnected. |
| 15. | ResourceReconnected | Hooks the event when resource get reconnected and also updates the list of available resources. |
| 16. | ResourceDisconnected | Called when the resource get disconnected and updates the list of available resources. |

Let’s take an example of FIFO strategy which is already implemented in the proposed API. It extends the AlgorithmBase class and provides an implementation of the First-In First-Out scheduling strategy. In this algorithm, the tasks are scheduled in their order of arrival. The Schedule function is changed in this implementation which checks that there is a task in the queue and a resource is also free. If both these conditions are true, then it schedules the first task in the queue to the first resource in the resource list. Similarly, other algorithms are proposed in the API and these algorithms are self-explained.

### 3.4 Aneka.Scheduling.Event

It provides all the events associated with the scheduling policies of Aneka. Proposed API provides the flexibility to end user to use already available events or create their own specific event. These events help scheduling policy designer to achieve desired performance and usability. These events are related to different components of Aneka which are listed in Table 4.

Table 4. Different type of events created in Aneka.Scheduling.Event

| **S. No.** | **Event Type** | **Description** |
|---|---|---|
| 1. | Task Events | These events are related to tasks such as task finished, task aborted etc. |
| 2. | Scheduling Events | These events are related to scheduling such as algorithm selected etc. |
| 4. | Resource Pool Query Events | These events are related adding, selecting and deleting specific resource pools. |
| 5. | Resource Events | These events are related to resource addition, deletion and selection. |

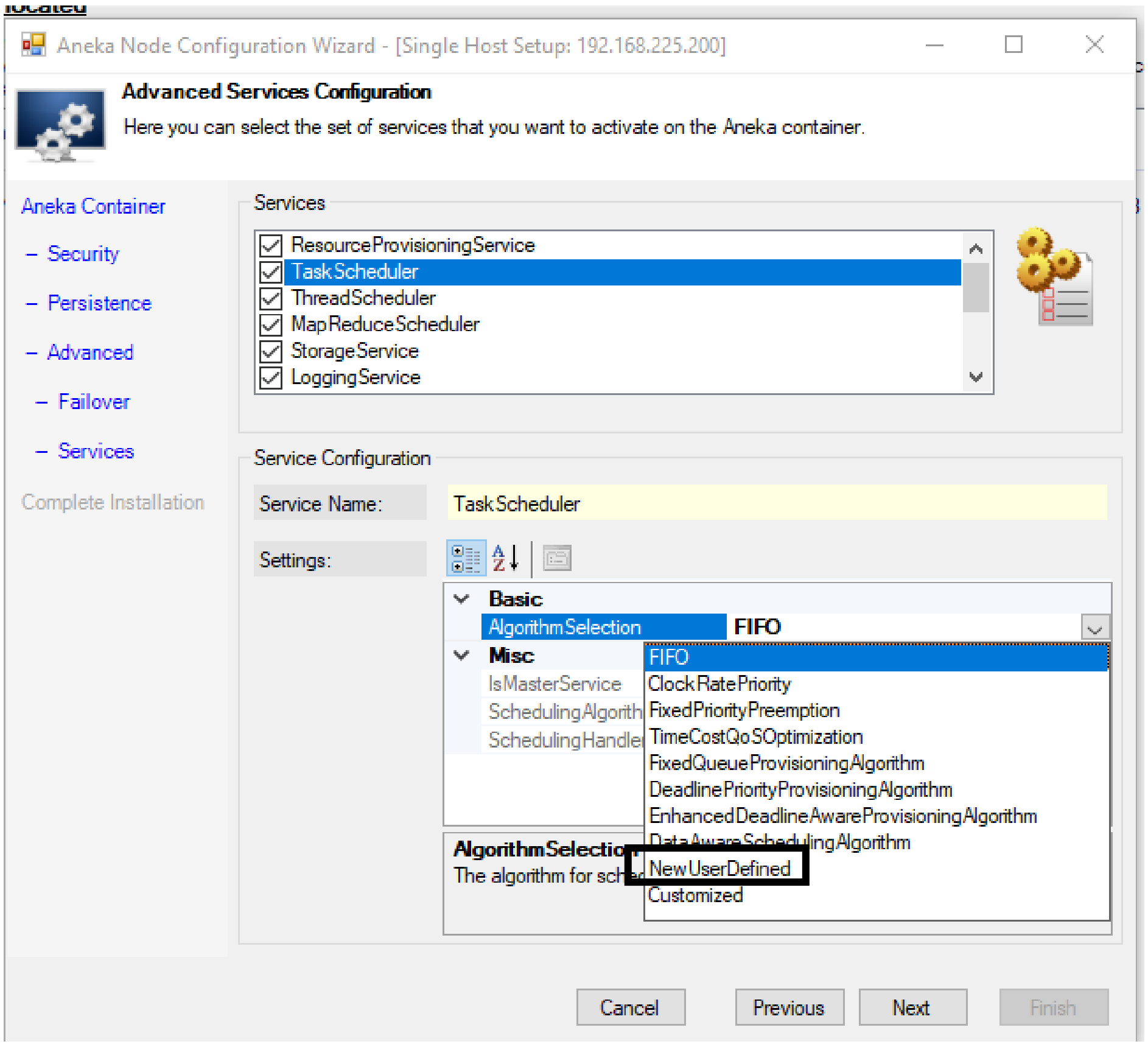

Figure 8 Interface in Management Studio to Select NewUserDefined Scheduling Algorithm

### 3.5 Selection of NewUserDefined from Management Studio

In this section step by step method to integrate the newly proposed scheduling policy is discussed. End user has to write new scheduling algorithm in NewUserDefined class of Aneka.Scheduling.Algorihtm. When Aneka master is being installed NewUserDefined should be chosen as shown in Figure 8. After this all the application running on Aneka if use dynamic provisioning will follow the scheduling policy desired in the proposed scheduling algorithm.

## 4. Developing New Scheduling Algorithms

Scheduling algorithms in Aneka define the logic with which tasks are allocated to resources from Aneka runtime. Developing new scheduling involves writing the code for the new algorithm and plug it into the existing scheduling services with minimal knowledge of the internals of the Aneka runtime. As we stated earlier this can be done by implementing methods of two interfaces — namely ISchedulerContext and ISchedulingAlgorithm — that represent the interface with the Aneka runtime and the scheduling algorithm respectively. These interfaces and bases classes are part of the Aneka.Scheduling library.

```
using Aneka.Provisioning;
using Aneka.Scheduling.Entity;
using Aneka.Scheduling.Event;

namespace Aneka.Scheduling
{
   public interface ISchedulerContext
   {
        bool SupportsProvisioning { get; }

        event EventHandler<SchedulingEventArgs> AssignTask;

        event EventHandler<ProvisionResourcesArgs> ProvisionResources;
```

```
        event EventHandler<ReleaseResourcesArgs> ReleaseResources;

        void Start();

        void Stop();

        void AddTasks(params Task[] tasks);

        void AddResources(params Resource[] resources);

        void SetScheduler(ISchedulerContext scheduler);
    }
}
```

Figure 9 *ISchedulingAlgorithm* interface

Figure 9 shows the ISchedulingAlgorithm interface that each scheduling algorithm has to implement. The algorithm provides a feedback to the Aneka runtime about its scheduling decisions through the events exposed by the interface. SupportsProvisioning is a boolean value that is set to true if the algorithm supports dynamic provisioning. AssignTask is trigred whenever a task is allocated to a resource. and ReleaseResources are events which are triggered when the scheduling algorithm issues a request for additional resources and reguest release of a provisioned resource, respectively. Start and Stop methods are called when the scheduling begins and ends. To add new tasks and new resources AddTasks and AddResources methods are called respectively. The algorithm works with the Aneka runtime and interfaces by means of the ISchedulerContext interface which is set in the SetScheduler method. Figure 10 shows ISchedulerContext interface.

```
using Aneka.Provisioning;
using Aneka.Scheduling.Entity;
using Aneka.Scheduling.Event;
using Aneka.Scheduling.Runtime;

namespace Aneka.Scheduling
{
    public interface ISchedulerContext
    {
        ISchedulingAlgorithm SchedulerAlgorithm { get; set; }

        ISchedulingHandler SchedulingHandler { get; set; }

        event EventHandler<ResourceEventArgs> ResourceDisconnected;

        event EventHandler<ResourceEventArgs> ResourceReConnected;

        event EventHandler<ProvisionEventArgs> ResourceProvisionProcessed;

        event EventHandler<ProvisionResourcesArgs> ResourceProvisionRequested;

        event EventHandler<ReleaseResourcesArgs> ResourceReleaseRequested;


        event EventHandler<TaskEventArgs> TaskFinished;

        event EventHandler<TaskEventArgs> TaskFailed;

        event EventHandler<TaskEventArgs> TaskAborted;

        event EventHandler<TaskEventArgs> TaskRequeued;
    }
}
```

Figure 9 *ISchedulingContext* interface

Only a subset of events and properties are of interest for the scheduling algorithm.

- *ResourceDisconnected* and *ResourceReconnected*: notify the scheduling algorithm that a resource has disconnected or reconnected from a temporary disconnection.
- *TaskAborted*, *TaskFinished*, *TaskFailed*, and *TaskRequeued*: notify the scheduling algorithm of the status of the tasks.
- *ResourceProvisionProcessed*: is the only event from dynamic provisioning infrastructure that is of interest for the scheduling algorithm. This event provides information about the outcome of a resource provisioning request made earlier by the scheduling algorithm.

### 4.1 Example 1: FIFO Scheduling Algorithm

As stated earlier. *AlgorithmBase* implements the *ISchedulingAlgorithm* interface and can be used as a template for creating specialized algorithms. Figure 11 provides *FIFOSchedulingAlgorithm* Class body to show by simply overriding a few methods of the *AlgorithmBase* Class First-In First-Out scheduling strategy can be implemented where the tasks are scheduled in their order of arrival.

```
using Aneka.Scheduling.Entity;
using Aneka.Scheduling.Event;
using Aneka;

namespace Aneka.Scheduling.Algorithms.Independent
{
public class FIFOSchedulingAlgorithm : AlgorithmBase
    {
     private List<Resource> _freeList = new List<Resource>();
     protected override bool HaveFreeResources()
        {
            return _freeList.Count > 0;
        }
     protected override void AddFreeResource(Resource r)
        {
            lock (this.synchLock)
            {
                int track = DebugUtil.EnterLock();
                if (r.IsConnected && r.FreeSlots > 0)
                {
                   _freeList.Remove(r);
                   _freeList.Add(r);
                }

                DebugUtil.ExitLock(track);
            }
        }
     protected override void RemoveFreeResource(Resource r)
        {
            lock (this.synchLock)
            {
                int track = DebugUtil.EnterLock();
                _freeList.Remove(r);

                DebugUtil.ExitLock(track);
            }
        }
     protected override void Schedule()
        {
            lock (this.synchLock)
            {
                int track = DebugUtil.EnterLock();

                if (_freeList.Count > 0 && TasksInQueue > 0)
                {
                   StartScheduleTask(_freeList.AsReadOnly(),GetNextTask());
                }
                else
                {
```

```
                    canSchedule.Reset();
                 }

                 DebugUtil.ExitLock(track);
             }
         }
     }
 }
```

Figure 11 *FIFOSchedulingAlgorithm Class*

*_freeList* maintains the list of the currently available resources for scheduling tasks. *HaveFreeResources()* method returns a Boolean value indicating whether there are free resources that can be used. *AddFreeResource* is called when a resource has a task removed. Since the recourse might be already on the free list, we remove it first and we add it again to avoid duplicates. *RemoveFreeResource* is called when adding a task to a resource fills all the slots or will be called for other reasons e.g., a disconnected resource. *Schedule()* starts the scheduling algorithm. If there is a task in the queue and a free resource in the *_freeList* it calls the *StartScheduleTask* method by passing *_freeList* and the next task in the queue. Otherwise, it resets the *canSchedule* to block the scheduling thread until the task can be scheduled again by any other resource.

### 4.2 Example 2: Deadline Priority Provisioning Algorithm

A new scheduling algorithm can be designed in a way that supports dynamic provisioning of virtual resources by leveraging the resource provisioning service. These are all defined in the namespace *Aneka.Scheduling.Algorithm.Independent*, which can be found in the Aneka library. *ProvisioningAlgorithmBase* is this class provides an abstract base class for all dynamic provisioning algorithms. The algorithm provides a basic management of the provisioning request that has been issued. *ApplicationAwareProvisioningAlgorithm* is specialized for the scheduling a collection of tasks as a whole in order to ensure that some specific QoS parameters that are defined for the application are met. Developers can design their new scheduling algorithms and new strategies for triggering resource provisioning by extending one of these two classes or specializing the previous two algorithms.

For example, Figure 12 shows *DeadlinePriorityProvisioningAlgorithm* that extends *ApplicationAwareProvisionAlgorithmBase* class and leverages dynamic provisioning in order to schedule the execution of the tasks within the expected deadline. If the local resources are not enough to execute all the tasks in time, a request for additional resources is issued. The class overrides two main methods of the base class called *ShrinkRequired* and *GrowRequired* to request release or adding of resources from the provisioner respectively. Both methods use the private method called ExceedResourceCapacity to set a Boolean indicator called required.

```
namespace Aneka.Scheduling.Algorithms.Independent
{
    public class DeadlinePriorityProvisioningAlgorithm : ApplicationAwareProvisionAlgorithmBase
    {
        protected override bool ShrinkRequired(Aneka.Scheduling.Entity.Task task, QoS qos)
        {
            bool required = false;
            if (qos != null)
            {
                int currentResources = this.GetResourceCount(task.ApplicationId);
                if (qos.WorkRemaining<currentResources || qos.WorkCompleted==qos.TotalWork)
                {
                       required = true;
                }
                else
                {
                    required = this.ExceedResourceCapacity(qos, task.ApplicationId, false) == false &&
currentResources > 1;
                }
            }
            return required;
        }
        protected override bool GrowRequired(Aneka.Scheduling.Entity.Task task, QoS qos)
        {
            bool required = false;
            string applicationId = task.ApplicationId;
            if (qos != null)
            {
```

```
                required = this.ExceedResourceCapacity(qos, applicationId, true);
            }
            return required;
        }
}
```

Figure 12 *DeadlinePriorityProvisioningAlgorithm Class*

The *ExceedResourceCapacity* method checks whether the current allocation for the application is compliant with the requirements set for the corresponding application. A Boolean value *toGrow* indicates whether we need to check for additional resources to add when it is *true* or resources to release when it is *false*. *taskRemaining* keeps the total number of remaining tasks that must be executed. *taskResourceRatio* is then calculated based on the ratio of the number of remaining tasks to the number of current resources. Finally based on the indicative values of *AverageTaskExecutionTime* in the *requiredTime* is calculated and is compared to *timeRemaining* which indicates the time remaining to the deadline. *bRequired* value is then set accordingly to true when the required time is larger than the remaining time.

```
private bool ExceedResourceCapacity(QoS qos, string applicationId, bool toGrow)
        {
            bool bRequired = false;
            int currentResources = this.GetResourceCount(applicationId);
            if (currentResources > 0)
            {
                int taskRemaining = 0;
                if (toGrow == true)
                {
                    taskRemaining = qos.TotalWork - qos.ScheduledTasks;
                }
                else
                {
                    taskRemaining = qos.TotalWork - qos.WorkCompleted;
                }
                int taskResourceRatio = taskRemaining / currentResources;
                TimeSpan avgExecutionTimeForTask = qos.AverageTaskExecutionTime;
                TimeSpan timeRemaining = qos.TimeRemaining;
                double requiredTime = avgExecutionTimeForTask.TotalSeconds * taskResourceRatio;
                if (requiredTime > timeRemaining.TotalSeconds)
                {
                    bRequired = true;
                }
            }
            else
            {
                bRequired = true;
            }
            return bRequired;
        }
```

Figure 12 *ExceedResourceCapacity* method

## 5. Performance Evaluation

For the sake of performance evaluation, we built a small-scale experimental testbed and tested our previously proposed resource provisioning and scheduling algorithm called Data-aware [12]. We use our proposed API all to integrate Data-aware scheduling algorithm. For more details on the Data-aware algorithm and workload setup, please look at our paper in [12]. The testbed is a hybrid cloud environment constituting of two desktop machines (one master and one slave) residing at The University of Melbourne and dynamic resources provisioned from Microsoft Azure. Configurations of resources used in the experiment are shown in Table 5. Public cloud resources are dynamically provisioned from Microsoft Azure cloud when local resources are not able to meet application deadlines.

Table 5: Configuration of machines used in the experiments.

| Machine | Type | CPU | Cores | Memory | OS |
|---|---|---|---|---|---|
| Master | Intel Core i7-4790 | 3.60 GHz | 8 | 16GB | Windows 7 |
| Worker | Intel Core i7-2600 | 3.40 GHz | 8 | 8GB | Windows 7 |
| Azure Instances | Standard DS1 | 2.4 GHz | 1 | 3.5GB | Windows Server 2012 |

As an application, a Bag-of-Tasks for measuring a walkability index is used [12]. A walkability index is used to assess how walkable a given neighborhood is based on factors such as road connectivity, gross dwelling density and the land use mix in the area. In our experiments, we use the walkability application to provide walkability indexes for 220 different neighborhoods in the city of Melbourne. The walkability application suits the purpose of our experiments since it is data-intensive and it can be broken into independent tasks, each computing a walkability index for a neighborhood. The test application contains 55 tasks, each calculating walkability indexes for four different neighborhoods of Melbourne city.

### 5.1 Experimental Results

We submit the walkability application to Aneka for execution with different deadlines, showing how the proposed algorithm behaves. All experiments are repeated for the other Aneka inbuilt scheduling algorithms, called *Default* and *Enhanced*. The Default scheduling algorithm makes an estimation of the expected completion time of the application with currently available resources, and if the expected completion time is later than the deadline defined in the Quality of Service parameters of the application, it requests extra resources from the public cloud to complete the application within given deadlines. The Enhanced algorithm is designed to utilize Amazon EC2 Spot Instance resources with an average higher deployment time but lower budget than the Default algorithm.
The execution time of the application without setting a deadline and only using private (local) resources takes 45.4 minutes. Figure 12 shows the results of the application execution under different deadlines. As shown by Figure 12, scheduling algorithms meet the deadline in all scenarios except in 3 cases, highlighted by “x” in the figure. Default and Enhanced algorithms violate the deadline constraint when the deadline is set to 35 minutes. The Default algorithm also misses the deadline when it is set to 40 minutes. The key reason is that these algorithms rely on only a single variable for measuring average runtime of tasks to allocate dynamic resources without considering data transfer time.

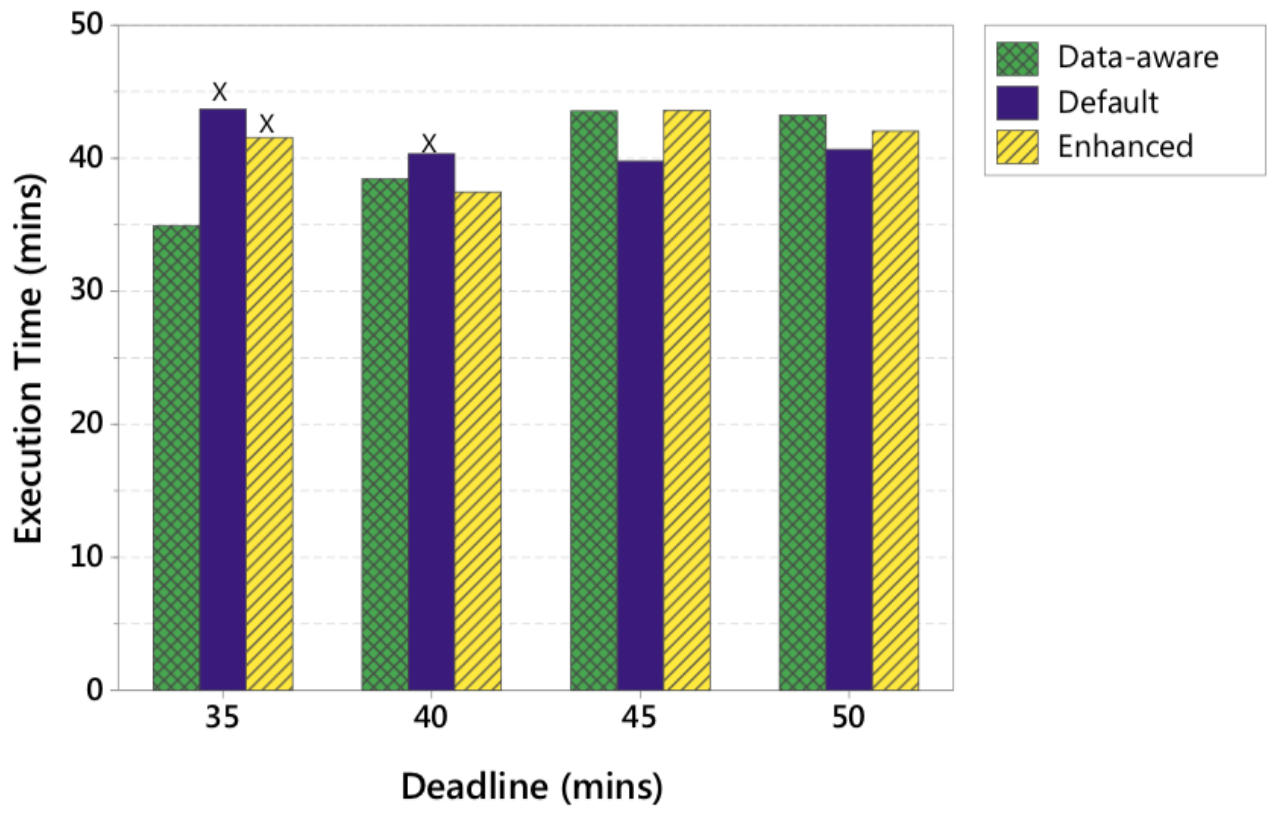

Figure 12 Execution time for Default, Enhanced, and Data-aware algorithms considering different application deadlines. The X symbol shows a violated deadline.

Experimental results demonstrate that the we can integrate new scheduling algorithm to Aneka Software using the proposed API. The results also show that the plugged-in scheduling algorithm works with qualitatively similar performance as the inbuilt scheduling algorithms and even outperforms them in some cases in terms of meeting deadline constraints.

## 6. Conclusion and Future Work

Aneka is one of the prominent PaaS cloud software available in the market which allows you to change and manage the underline infrastructure as well as write applications in any programming model. Aneka has many features the research community well demands the customization of Aneka scheduling algorithm for dynamic provisioning. In this chapter, an API is proposed which provides all necessary libraries to develop new Aneka scheduling algorithm. The new algorithm created from this API can be easily integrated with Aneka interface using Management Studio. The proposed API will help cloud computing researchers to develop their application in a real cloud and test them with their custom scheduling algorithm.

As part of the future work, we will focus on the creation of graphical user interface for addition of new scheduling APIs. This interactive interface will feature the designing of scheduling algorithms for new paradigms such as Internet of Things (IoT) and Fog Computing. The future work will also consist of development of APIs which are compatible with creation of multi-level scheduling policies such as from IoT to Fog layer to Cloud computing layer. Using multi-layer scheduling policies users can design scheduling policies for latency sensitive applications running in Fog computing environments.

**References**

[1] R. Buyya, "Market-oriented cloud computing: Vision, hype, and reality of delivering computing as the 5th utility," in *2009 9th IEEE/ACM International Symposium on Cluster Computing and the Grid, CCGRID 2009*, 2009.

[2] J. Gubbi, R. Buyya, S. Marusic, and M. Palaniswami, "Internet of Things (IoT): A vision, architectural elements, and future directions," *Futur. Gener. Comput. Syst.*, vol. 29, no. 7, pp. 1645–1660, Sep. 2013.

[3] W. J. Wang, Y. S. Chang, W. T. Lo, and Y. K. Lee, "Adaptive scheduling for parallel tasks with QoS satisfaction for hybrid cloud environments," *J. Supercomput.*, vol. 66, no. 2, pp. 783–811, Feb. 2013.

[4] R. N. Calheiros, R. Ranjan, A. Beloglazov, C. A. F. De Rose, and R. Buyya, "CloudSim: A toolkit for modeling and simulation of cloud computing environments and evaluation of resource provisioning algorithms," *Softw. - Pract. Exp.*, vol. 41, no. 1, pp. 23–50, Jan. 2011.

[5] H. Gupta, A. Vahid Dastjerdi, S. K. Ghosh, and R. Buyya, "iFogSim: A toolkit for modeling and simulation of resource management techniques in the Internet of Things, Edge and Fog computing environments," *Softw. - Pract. Exp.*, vol. 47, no. 9, pp. 1275–1296, Sep. 2017.

[6] X. Zeng, S. K. Garg, P. Strazdins, P. P. Jayaraman, D. Georgakopoulos, and R. Ranjan, "IOTSim: A simulator for analysing IoT applications," *J. Syst. Archit.*, 2017.

[7] "AWS - Amazon EC2 Instance Types," *Amazon*, 2014. [Online]. Available: http://aws.amazon.com/ec2/instance-types/. [Accessed: 19-Jul-2015].

[8] Microsoft Azure, "Microsoft Azure Pricing calculator," *Microsoft*, 2016. [Online]. Available: https://azure.microsoft.com/en-us/pricing/calculator/.

[9] T. Shon, J. Cho, K. Han, and H. Choi, "Toward advanced mobile cloud computing for the internet of things: Current issues and future direction," *Mob. Networks Appl.*, vol. 19, no. 3, pp. 404–413, Jun. 2014.

[10] C. Vecchiola, X. Chu, and R. Buyya, "Aneka: a software platform for .NET-based cloud computing," *High Speed Large Scale Sci. Comput.*, 2009.

[11] R. Buyya and D. Barreto, "Multi-cloud resource provisioning with Aneka: A unified and integrated utilisation of microsoft azure and amazon EC2 instances," in *2015 International Conference on Computing and Network Communications, CoCoNet 2015*, 2016.

[12] A. Nadjaran Toosi, R. O. Sinnott, and R. Buyya, "Resource provisioning for data-intensive applications with deadline constraints on hybrid clouds using Aneka," *Futur. Gener. Comput. Syst.*, 2018.